\documentclass[12pt,english]{article}
\usepackage[T1]{fontenc}
\usepackage{lmodern}
\usepackage[latin9]{inputenc}
\usepackage{geometry}
\usepackage{color}
\usepackage{float}
\usepackage{amsmath}
\usepackage{graphicx}
\usepackage{setspace}
\usepackage{esint}
\usepackage{babel}
\begin{document}

\title{\textbf{Coupled mass-momenta balance for modeling material failure}}

\author{\textbf{K.Y. Volokh}\thanks{E-mail: cvolokh@technion.ac.il}\\
\textit{Faculty of Civil and Environmental Engineering, Technion -
I.I.T., Israel}}
\maketitle
\begin{abstract}
Cracks are created by massive breakage of molecular or atomic bonds.
The latter, in its turn, leads to the highly localized loss of material,
which is the reason why even closed cracks are visible by a naked
eye. Thus, fracture can be interpreted as the local material sink.
Mass conservation is violated locally in the area of material failure.
We consider a theoretical formulation of the coupled mass and momenta
balance equations for a description of fracture. Our focus is on brittle
fracture and we propose a finite strain hyperelastic thermodynamic
framework for the coupled mass-flow-elastic boundary value problem.
The attractiveness of the proposed framework as compared to the traditional
continuum damage theories is that no internal parameters (like damage
variables, phase fields etc.) are used while the regularization of
the failure localization is provided by the physically sound law of
mass balance. 
\end{abstract}

\section{Introduction}

Within the framework of continuum mechanics there are surface and
bulk material failure models.

Surface failure models are known by name of Cohesive Zone Models (CZMs).
In the latter case, continuum is enriched with discontinuities along
surfaces - cohesive zones - with additional traction-displacement-separation
constitutive laws. These laws are built qualitatively as follows:
traction increases up to a maximum and then goes down to zero via
increasing separation (Barenblatt, 1959; Needleman, 1987; Rice and
Wang, 1989, Tvergaard and Hutchinson, 1992; Camacho and Ortiz, 1996;
de Borst, 2001; Xu and Needleman, 1994; Roe and Siegmund, 2003; Moes
et al, 1999; Park et al, 2009; Gong et al, 2012). If the location
of the separation surface is known in advance (e.g. fracture along
weak interfaces) then the use of CZM is natural. Otherwise, the insertion
of cracks in the bulk in the form of the separation surfaces remains
an open problem, which includes definition of the criteria for crack
nucleation, orientation, branching and arrest. Besides, the CZM approach
presumes the simultaneous use of two different constitutive models,
one for the cohesive zone and another for the bulk, for the same real
material. Certainly, a correspondence between these two constitutive
theories is desirable yet not promptly accessible. The issues concerning
the CZM approach have been discussed by Needleman (2014), the pioneer
of the field. 

Bulk failure models are known by name of Continuum Damage Mechanics
(CDM). In the latter case, material failure or damage is described
by constitutive laws including softening in the form of the falling
stress-strain curves (Kachanov, 1958; Gurson, 1977; Simo, 1987; Voyiadjis
and Kattan, 1992; Gao and Klein, 1998; Klein and Gao, 1998; Menzel
and Steinmann, 2001; Dorfmann and Ogden, 2004; Lemaitre and Desmorat,
2005; Volokh, 2004, 2007; Benzerga et al, 2016). Remarkably, damage
nucleation, propagation, branching and arrest naturally come out of
the constitutive laws. Unfortunately, numerical simulations based
on the the bulk failure laws show the so-called pathological mesh-sensitivity,
which means that the finer meshes lead to the narrower damage localization
areas. In the limit case, the energy dissipation in failure tends
to zero with the diminishing size of the computational mesh. This
physically unacceptable mesh-sensitivity is caused by the lack of
a characteristic length in the traditional formulation of continuum
mechanics. To surmount the latter pitfall gradient- or integral- type
nonlocal continuum formulations are used where a characteristic length
is incorporated to limit the size of the spatial failure localization
(Pijaudier-Cabot and Bazant, 1987; Lasry and Belytschko, 1988; Peerlings
et al, 1996; de Borst and van der Giessen, 1998; Francfort and Marigo,
1998; Silling, 2000; Hofacker and Miehe, 2012; Borden et al, 2012).
The regularization strategy rooted in the nonlocal continua formulations
is attractive because it is lucid mathematically.

Unluckily, the generalized nonlocal continua theories are based (often
tacitly) on the physical assumption of long-range particle interactions
while the actual particle interactions are short-range - on nanometer
or angstrom scale. Therefore, the physical basis for the nonlocal
models appears disputable. A more physically-based treatment of the
pathological mesh-sensitivity of the bulk failure simulations should
likely include multi-physics coupling. Such an attempt to couple mass
flow (sink) and finite elastic deformation within the framework of
brittle fracture is considered in the present work.

\section{Basic idea}

\begin{figure}[H]
\noindent \begin{centering}
\includegraphics{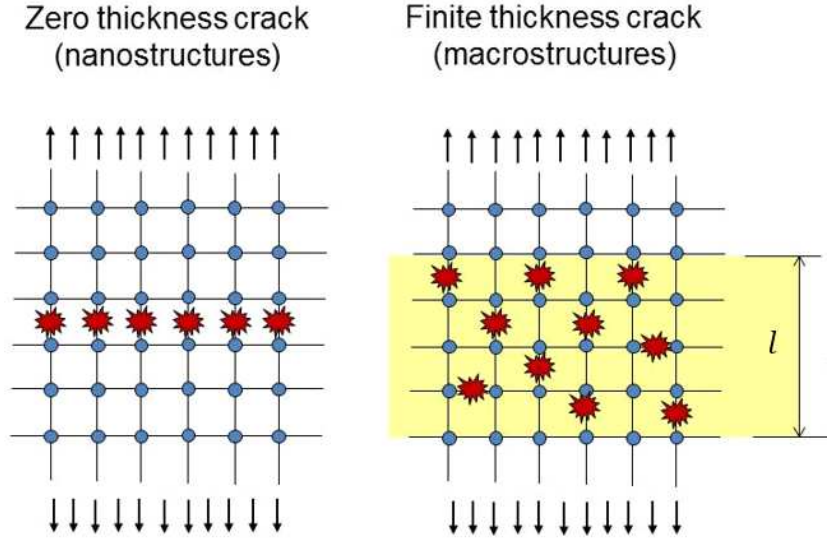}
\par\end{centering}

\caption{Schematic drawing of cracks with zero or finite thickness.\label{fig:Schematic-cracks-of}}
\end{figure}

Cracks are often thought of as material discontinuities of zero thickness.
Such idealized point of view is probably applicable to nano-structures
with perfect crystal organization. In the latter case fracture appears
as a result of a separation - unzipping - of two adjacent atomic or
molecular layers - Fig. \ref{fig:Schematic-cracks-of} (left). 

\noindent 

In the case of the bulk material with a sophisticated heterogeneous
organization the crack appears as a result of the development of multiple
micro-cracks triggered by the massive breakage of molecular or atomic
bonds - Fig. \ref{fig:Schematic-cracks-of} (right). The bond breakage
is not confined to two adjacent molecular layers and the process involves
thousands layers within an area or volume with the representative
characteristic size $l$.

It is in interesting that material failure does not require the breakage
of all molecular or atomic bonds within a representative volume. Only
fraction of these bonds should be broken for the material disintegration.
For example, in the case of natural rubber, roughly speaking, every
third bond should be broken within a representative volume to create
crack (Volokh, 2013a).

The local bond failure leads to the highly localized loss of material.
The latter, in our opinion, is the reason why even closed cracks are
visible by a naked eye. Thus, material flows out of the system during
the fracture process. The system becomes open from the thermodynamic
standpoint. However, cracks usually have very small thickness and
the amount of the lost material is negligible as compared to the whole
bulk. The latter observation prompts considering the system as the
classical closed one. Such approximation allows ignoring the additional
supply of momenta and energy in the formulation of the initial boundary
value problem described in the next sections.

\section{Thermodynamics}

Following the approach of continuum mechanics we replace the discrete
molecular structure of materials by a continuously distributed set
of material points which undergo mappings from the initial (reference),
$\varOmega_{0}$, to current, $\varOmega$, configuration: $\mathbf{x}\mapsto\mathbf{y}(\mathbf{x})$.
The deformation in the vicinity of the material points is described
by the deformation gradient $\mathbf{F}=\mathrm{Grad}\mathbf{y}(\mathbf{x})$. 

In what follows we use the Lagrangean description with respect to
the initial or reference configuration and define the local mass balance
in the form
\begin{equation}
\frac{d\rho}{dt}=\mathrm{Div}\mathbf{s}+\xi,
\end{equation}
where $\rho$ is the referential (Lagrangean) mass density; $\mathbf{s}$
is the referential mass flux; $\xi$ is the referential mass source
(sink); and $\mathrm{Div}\mathbf{s}=\partial s_{i}/\partial x_{i}$
in Cartesian coordinates.

\textit{We further assume that failure and, consequently, mass flow
are highly localized and the momenta and energy balance equations
can be written in the standard form without adding momenta and energy
due to the mass alterations.}

In view of the assumption above, we write momenta and energy balance
equations in the following forms accordingly
\begin{equation}
\frac{d(\rho\mathbf{v})}{dt}=\mathrm{Div}\mathbf{P}+\rho\mathbf{b},\quad\mathbf{P}\mathbf{F}^{\mathrm{T}}=\mathbf{F}\mathbf{P}^{\mathrm{T}},
\end{equation}
and
\begin{equation}
\frac{d(\rho e)}{dt}=\mathbf{P}:\dot{\mathbf{F}}+\rho r-\mathrm{Div}\mathbf{q},\label{eq:energy balance}
\end{equation}
where $\mathbf{v}=\dot{\mathbf{y}}$ is the velocity of a material
point; $\mathbf{b}$ is the body force per unit mass; $\mathbf{P}$
is the first Piola-Kirchhoff stress and $(\mathrm{Div}\mathbf{P})_{i}=\partial P_{ij}/\partial x_{j}$;
$e$ is the specific internal energy per unit mass; $r$ is the specific
heat source per unit mass; and $\mathbf{q}$ is the referential heat
flux.

Entropy inequality reads
\begin{equation}
\frac{d(\rho\eta)}{dt}\geq\frac{1}{T}(\rho r-\mathrm{Div}\mathbf{q})+\frac{1}{T^{2}}\mathbf{q}\cdot\mathrm{Grad}T,\label{eq:entropy inequality}
\end{equation}
where $T$ is the absolute temperature.

Substitution of $(\rho r-\mathrm{Div}\mathbf{q})$ from (\ref{eq:energy balance})
to (\ref{eq:entropy inequality}) yields
\begin{equation}
\rho\dot{\eta}+\dot{\rho}\eta\geq\frac{1}{T}(\rho\dot{e}+\dot{\rho}e-\mathbf{P}:\dot{\mathbf{F}})+\frac{1}{T^{2}}\mathbf{q}\cdot\mathrm{Grad}T,
\end{equation}
or, written in terms of the internal dissipation,
\begin{equation}
D_{\mathrm{int}}=\mathbf{P}:\dot{\mathbf{F}}-\rho(\dot{e}-T\dot{\eta})-\dot{\rho}(e-T\eta)-\frac{1}{T}\mathbf{q}\cdot\mathrm{Grad}T\geq0.\label{eq:dissipation 1}
\end{equation}

We introduce the specific Helmholtz free energy per unit mass
\begin{equation}
w=e-T\eta,
\end{equation}
and, consequently, we have
\begin{equation}
e=w+T\eta,\quad\dot{e}=\dot{w}+\dot{T}\eta+T\dot{\eta}.\label{eq:specific internal energy}
\end{equation}

Substituting (\ref{eq:specific internal energy}) in (\ref{eq:dissipation 1})
we get
\begin{equation}
D_{\mathrm{int}}=\mathbf{P}:\dot{\mathbf{F}}-\rho(\dot{w}+\dot{T}\eta)-\dot{\rho}w-\frac{1}{T}\mathbf{q}\cdot\mathrm{Grad}T\geq0.\label{eq:dissipation 2}
\end{equation}

Then, we calculate the Helmholtz free energy increment
\begin{equation}
\dot{w}=\frac{\partial w}{\partial\mathbf{F}}:\dot{\mathbf{F}}+\frac{\partial w}{\partial T}\dot{T},
\end{equation}
and substitute it in (\ref{eq:dissipation 2}) as follows
\begin{equation}
D_{\mathrm{int}}=(\mathbf{P}-\rho\frac{\partial w}{\partial\mathbf{F}}):\dot{\mathbf{F}}-\rho(\frac{\partial w}{\partial T}+\eta)\dot{T}-\dot{\rho}w-\frac{1}{T}\mathbf{q}\cdot\mathrm{Grad}T\geq0.
\end{equation}

The Coleman-Noll procedure suggests the following choice of the constitutive
laws
\begin{equation}
\mathbf{P}=\rho\frac{\partial w}{\partial\mathbf{F}},\quad\eta=-\frac{\partial w}{\partial T}.
\end{equation}
and, consequently, the dissipation inequality reduces to
\begin{equation}
D_{\mathrm{int}}=-\dot{\rho}w-\frac{1}{T}\mathbf{q}\cdot\mathrm{Grad}T\geq0.\label{eq:dissipation 3}
\end{equation}

\textit{We further note that the process of the bond breakage is very
fast as compared to the dynamic deformation process and the mass density
changes in time as a step function. So, strictly speaking, the density
rate should be presented by the Dirac delta in time. We will not consider
the super fast transition to failure, which is of no interest on its
own, and assume that the densities before and after failure are constants
and, consequently,} 
\begin{equation}
\dot{\rho}=\mathrm{Div}\mathbf{s}+\xi=0.\label{eq:mass flow}
\end{equation}

Then, the dissipation inequality reduces to
\begin{equation}
D_{\mathrm{int}}=-\frac{1}{T}\mathbf{q}\cdot\mathrm{Grad}T\geq0,\label{eq:dissipation 4}
\end{equation}
which is obeyed because the heat flows in the direction of the lower
temperature.

It remains to settle the boundary and initial conditions.

Natural boundary conditions for zero mass flux represent the mass
balance on the boundary $\partial\varOmega_{0}$
\begin{equation}
\mathbf{s}\cdot\mathbf{n}=0,
\end{equation}
or
\begin{equation}
\mathbf{n}\cdot\mathrm{Grad}\rho=0,
\end{equation}
where $\mathbf{n}$ is the unit outward normal to the boundary in
the reference configuration.

Natural boundary conditions for given traction $\bar{\mathbf{t}}$
represent the linear momentum balance on the boundary $\partial\varOmega_{0}$
\begin{equation}
\mathbf{P}\mathbf{n}=\bar{\mathbf{t}},
\end{equation}
or, alternatively, the essential boundary conditions for placements
can be prescribed on $\partial\varOmega_{0}$
\begin{equation}
\mathbf{y}=\bar{\mathbf{y}}.
\end{equation}

Initial conditions in $\varOmega_{0}$ complete the formulation of
the coupled mass-flow-elastic initial boundary value problem
\begin{equation}
\mathbf{y}(t=0)=\mathbf{y}_{0},\quad\mathbf{v}(t=0)=\mathbf{v}_{0}.
\end{equation}

\section{Constitutive equations}

Constitutive law for the Lagrangean mass flux can be written by analogy
with the Fourier law for heat conduction
\begin{equation}
\mathbf{s}=\kappa J(\mathbf{F}^{\mathrm{T}}\mathbf{F})^{-1}\mathrm{Grad}\rho,\label{eq:mass flux}
\end{equation}
where $\kappa>0$ is a mass conductivity constant for the isotropic
case.

Constitutive law for the mass source is the very heart of the successful
formulation of the theory and the reader is welcome to make a proposal.

We choose, for example, the following constitutive law, whose motivation
is clarified below,
\begin{equation}
\xi(\rho,\rho_{0},w,\phi)=\beta J^{-1}(\rho_{0}H(\zeta)\exp[-(w/\phi)^{m}]-\rho),\label{eq:mass source}
\end{equation}
where $\rho_{0}=\rho(t=0)$ is a constant initial density; $\beta>0$
is a material constant; $\phi$ is the specific energy limiter per
unit mass, which is calibrated in macroscopic experiments; $m$ is
a dimensionless material parameter, which controls the sharpness of
the transition to material failure on the stress-strain curve; and
$H(\zeta)$ is a unit step function, i.e. $H(\zeta)=0$ if $\zeta<0$
and $H(\zeta)=1$ otherwise.

The switch parameter $\zeta$, which is necessary to prevent from
material healing, will be explained below.

Substitution of (\ref{eq:mass source}) and (\ref{eq:mass flux})
in (\ref{eq:mass flow}) yields
\begin{equation}
l^{2}\mathrm{Div}(J(\mathbf{F}^{\mathrm{T}}\mathbf{F})^{-1}\mathrm{Grad}\frac{\rho}{\rho_{0}})+H(\zeta)J^{-1}\exp[-(w/\phi)^{m}]-\frac{\rho}{J\rho_{0}}=0,\label{eq:mass balance}
\end{equation}
where
\begin{equation}
l=\sqrt{\kappa/\beta}
\end{equation}
is the characteristic length.

\textit{It is remarkable that we, actually, do not need to know $\kappa$
and $\beta$ separately and the knowledge of the characteristic length
is enough}. For example, the estimate of the characteristic length
for rubber is $l=0.2\:\mathrm{mm}$ (Volokh, 2011) and for concrete
it is $l=2.6\:\mathrm{cm}$ (Volokh, 2013b).

To justify the choice of the constitutive equation (\ref{eq:mass source})
for the mass source/sink we note that in the case of the homogeneous
deformation and mass flow the first term on the left hand side of
(\ref{eq:mass balance}) vanishes and we obtain
\begin{equation}
\rho=\rho_{0}H(\zeta)\exp[-(w/\phi)^{m}].
\end{equation}

Substituting this mass density in the hyperelastic constitutive law
we have
\begin{equation}
\mathbf{P}=\rho_{0}H(\zeta)\exp[-(w/\phi)^{m}]\frac{\partial w}{\partial\mathbf{F}}=H(\zeta)\exp[-(W/\varPhi)^{m}]\frac{\partial W}{\partial\mathbf{F}},\label{eq:stress-strain}
\end{equation}
where
\begin{equation}
W=\rho_{0}w,\quad\varPhi=\rho_{0}\phi
\end{equation}
are the Helmholtz free energy and energy limiter per unit referential
volume accordingly.

Constitutive law (\ref{eq:stress-strain}) presents the hyperelasticity
with the energy limiters - see Volokh (2007, 2013a, 2016) for the
general background. Integrating (\ref{eq:stress-strain}) with respect
to the deformation gradient we introduce the following form of the
strain energy function
\begin{equation}
\varPsi(\mathbf{F},\zeta)=\varPsi_{\mathrm{f}}-H(\zeta)\varPsi_{\mathrm{e}}(\mathbf{F}),\label{eq:energy with limiter}
\end{equation}
where
\begin{equation}
\varPsi_{\mathrm{e}}(\mathbf{F})=\frac{\varPhi}{m}\varGamma(\frac{1}{m},\frac{W(\mathbf{F})^{m}}{\varPhi^{m}}),\quad\varPsi_{\mathrm{f}}=\varPsi_{\mathrm{e}}(\mathbf{1}).
\end{equation}

Here $\varPsi_{\mathrm{f}}$ and $\varPsi_{\mathrm{e}}(\mathbf{F})$
designate the constant bulk failure energy and the elastic energy
respectively; $\varGamma(s,x)=\int_{x}^{\infty}t^{s-1}e^{-t}dt$ is
the upper incomplete gamma function.

The switch parameter $\zeta\in(-\infty,0]$ is defined by the evolution
equation
\begin{equation}
\dot{\zeta}=-H(\epsilon-\frac{\varPsi_{\mathrm{e}}}{\varPsi_{\mathrm{f}}}),\quad\zeta(t=0)=0,\label{eq:switch parameter evolution}
\end{equation}
where $0<\epsilon\ll1$ is a dimensionless precision constant.

The physical interpretation of (\ref{eq:energy with limiter}) is
straight: material is hyperelastic for the strain energy below the
failure limit - $\varPsi_{\mathrm{f}}$. When the failure limit is
reached, then the strain energy becomes constant for the rest of the
deformation process precluding the material healing. Parameter $\zeta\leq0$
is \textit{not an internal variable}. It is a switch: $\zeta=0$ for
the reversible process; and $\zeta<0$ for the irreversibly failed
material and dissipated strain energy.

For illustration, we present the following specialization of the intact
strain energy for a filled Natural Rubber (NR) (Volokh, 2010)
\begin{equation}
{\color{black}W=\rho_{0}w=\sum_{k=1}^{3}c_{k}(I_{1}-3)^{k},\quad J}=\det\mathbf{F}=1,
\end{equation}
where $c_{1}=0.298\:\mathrm{MPa}$, $c_{2}=0.014\:\mathrm{MPa}$,
$c_{3}=0.00016\:\mathrm{MPa}$ and the failure parameters are $m=10$,
and $\varPhi=82.0\,\textrm{MPa}$.

\begin{figure}[H]
\begin{centering}
\includegraphics{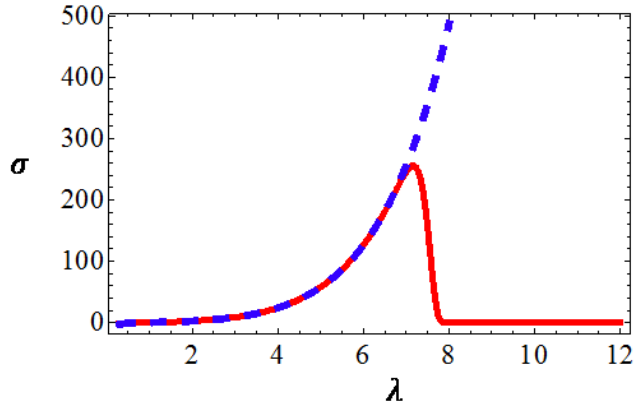}
\par\end{centering}

\caption{Uniaxial tension of natural rubber: Cauchy stress {[}MPa{]} versus
stretch. Dashed line specifies the intact model; solid line specifies
the model with energy limiter.\label{fig:Cauchy-stress}}
\end{figure}

The Cauchy stress, defined by $\boldsymbol{\sigma}=J^{-1}\mathbf{P}\mathbf{F}^{\mathrm{T}}$,
versus stretch curve for the uniaxial tension is shown in Fig. \ref{fig:Cauchy-stress}
for both cases with and without the energy limiter. Material failure
takes place at the critical limit point in correspondence with tests
conducted by Hamdi et al (2006).

For the implications and experimental comparisons of the elasticity
with energy limiters the reader is advised to look through Volokh
(2013a; 2016), for example. We completely skip this part for the sake
of brevity.

Thus, the proposed constitutive law for the mass source is motivated
by the limit case of the coupled formulations in which the deformation
is homogeneous.

\section{Conclusion}

Crack in a bulk material is not an ideal unzipping of two adjacent
atomic layers. It is rather a massive breakage of atomic bonds diffused
in a volume of characteristic size. The massive bond breakage is accompanied
by the localized loss of material. Thus, material sinks in the vicinity
of the crack. Evidently, the law of mass conservation should be replaced
by the law of mass balance, accounting for the mass flow in the vicinity
of the crack. The coupled mass-flow-elasticity problem should be set
for analysis of crack propagation.

In the present work, we formulated the coupled problem based on the
thermodynamic reasoning. We assumed that the mass loss related to
the crack development was small as compared to the mass of the whole
body. In addition, we assumed that the process of the bond breakage
was very fast and the mass density jumped from the intact to failed
material abruptly allowing to ignore the transient process of the
failure development. These physically reasonable assumptions helped
us to formulate a simple coupled initial boundary value problem. \textit{In
the absence of failure localization into cracks the theory is essentially
the hyperelasticity with the energy limiters. However, when the failure
starts localizing into cracks the diffusive material sink activates
via the mass balance equation and it provides the regularization of
numerical simulations. }The latter regularization is due to the mass
diffusion - first term on the left hand side of (\ref{eq:mass balance}).

The attractiveness of the proposed framework as compared to the traditional
continuum damage theories is that no internal parameters (like damage
variables, phase fields etc.) are used while the regularization of
the failure localization is provided by the physically sound law of
mass balance.

A numerical integration procedure for the formulated coupled initial
boundary value problem is required and it will be considered elsewhere.

\section*{Acknowledgement}

The support from the Israel Science Foundation (ISF-198/15) is gratefully
acknowledged.

\section*{References}

\noindent Barenblatt GI (1959) The formation of equilibrium cracks
during brittle fracture. General ideas and hypotheses. Axially-symmetric
cracks. J Appl Math Mech 23:622-636\\
\\
Benzerga AA, Leblond JB, Needleman A, Tvergaard V (2016) Ductile failure
modeling. Int J Fract 201:29-80\\
\\
Borden MJ, Verhoosel CV, Scott MA, Hughes TJR, Landis CM (2012) A
phase-field description of dynamic brittle fracture. Comp Meth Appl
Mech Eng 217-220:77-95\\
\\
de Borst R (2001) Some recent issues in computational failure mechanics.
Int J Numer Meth Eng 52:63-95\\
\\
de Borst R, van der Giessen E (1998) Material Instabilities in Solids.
John Wiley \& Sons, Chichester\\
\\
Camacho GT, Ortiz M (1996) Computational modeling of impact damage
in brittle materials. Int J Solids Struct 33:2899 \textendash 2938\\
\\
Dorfmann A, Ogden RW (2004) A constitutive model for the Mullins effect
with permanent set in particle-reinforced rubber. Int J Solids Struct
41:1855-1878\\
\\
Francfort GA, Marigo JJ (1998) Revisiting brittle fracture as an energy
minimization problem. J Mech Phys Solids 46:1319-1342\\
\\
Gao H, Klein P (1998) Numerical simulation of crack growth in an isotropic
solid with randomized internal cohesive bonds. J Mech Phys Solids
46:187-218\\
\\
Gong B, Paggi M, Carpinteri A (2012) A cohesive crack model coupled
with damage for interface fatigue problems. Int J Fract 137:91-104\\
\\
Gurson AL (1977) Continuum theory of ductile rupture by void nucleation
and growth: part I-yield criteria and flow rules for porous ductile
media. J Eng Mat Tech 99:2\textendash 151\\
\\
Hamdi A, Nait Abdelaziz M, Ait Hocine N, Heuillet P, Benseddiq N (2006)
A fracture criterion of rubber-like materials under plane stress conditions.
Polymer Testing 25:994-1005\\
\\
Hofacker M, Miehe C (2012) Continuum phase field modeling of dynamic
fracture: variational principles and staggered FE implementation.
Int J Fract 178:113-129\\
\\
Kachanov LM (1958) Time of the rupture process under creep conditions.
Izvestiia Akademii Nauk SSSR, Otdelenie Teckhnicheskikh Nauk 8:26-31\\
\\
Klein P, Gao H (1998) Crack nucleation and growth as strain localization
in a virtual-bond continuum. Eng Fract Mech 61:21-48\\
\\
Lasry D, Belytschko T (1988) Localization limiters in transient problems.
Int J Solids Struct 24:581-597\\
\\
Lemaitre J, Desmorat R (2005) Engineering Damage Mechanics: Ductile,
Creep, Fatigue and Brittle Failures. Springer, Berlin\\
\\
Menzel A, Steinmann P (2001) A theoretical and computational framework
for anisotropic continuum damage mechanics at large strains. Int J
Solids Struct 38:9505-9523\\
\\
Moes N, Dolbow J, Belytschko T (1999) A finite element method for
crack without remeshing. Int J Num Meth Eng 46:131-150\\
\\
Needleman A (1987) A continuum model for void nucleation by inclusion
debonding. J Appl Mech 54:525-531\\
\\
Needleman A (2014) Some issues in cohesive surface modeling. Procedia
IUTAM 10:221-246\\
\\
Park K, Paulino GH, Roesler JR (2009) A unified potential-based cohesive
model of mixed-mode fracture. J Mech Phys Solids 57:891-908\\
\\
Peerlings RHJ, de Borst R, Brekelmans WAM, de Vree JHP (1996) Gradient
enhanced damage for quasi-brittle materials. Int J Num Meth Eng 39:3391-3403\\
\\
Pijaudier-Cabot G, Bazant ZP (1987) Nonlocal damage theory. J Eng
Mech 113:1512-1533\\
\\
Rice JR, Wang JS (1989) Embrittlement of interfaces by solute segregation.
Mater Sci Eng A 107:23-40\\
\\
Roe KL, Siegmund T (2003) An irreversible cohesive zone model for
interface fatigue crack growth simulation. Eng Fract Mech 70:209-232\\
\\
Silling SA (2000) Reformulation of elasticity theory for discontinuities
and long-range forces. J Mech Phys Solids 48:175-209\\
\\
Simo JC (1987) On a fully three-dimensional finite strain viscoelastic
damage model: Formulation and computational aspects. Comp Meth Appl
Mech Eng 60:153-173\\
\\
Tvergaard V, Hutchinson JW (1992) The relation between crack growth
resistance and fracture process parameters in elastic-plastic solids.
J Mech Phys Solids 40:1377-1397\\
\\
Voyiadjis GZ, Kattan PI (1992) A plasticity-damage theory for large
deformation of solids\textemdash I. Theoretical formulation. Int J
Eng Sci 30:1089-1108\\
\\
Volokh KY (2004) Nonlinear elasticity for modeling fracture of isotropic
brittle solids. J Appl Mech 71:141-143\\
\\
Volokh KY (2007) Hyperelasticity with softening for modeling materials
failure. J Mech Phys Solids 55:2237-2264 \\
\\
Volokh KY (2010) On modeling failure of rubber-like materials. Mech
Res Com 37:684-689\\
\\
Volokh KY (2011) Characteristic length of damage localization in rubber.
Int J Fract 168:113-116\\
\\
Volokh KY (2013a) Review of the energy limiters approach to modeling
failure of rubber. Rubber Chem Technol 86:470-487\\
\\
Volokh KY (2013b) Characteristic length of damage localization in
concrete. Mech Res Commun 51:29-31\\
\\
Volokh KY (2016) Mechanics of Soft Materials. Springer\\
\\
Xu XP, Needleman A (1994) Numerical simulations of fast crack growth
in brittle solids. J Mech Phys Solids 42:1397-1434\\

\end{document}